\documentclass[twocolumn,superscriptaddress,secnumarabic,amssymb, nobibnotes, aps, bm,prl,floatfix,longbibliography]{revtex4-2}
\usepackage[english]{babel}
\usepackage[latin9]{inputenc}
\usepackage[usenames,dvipsnames]{xcolor}
\usepackage{mathtools}
\usepackage{amsmath}
\definecolor{goodgreen}{rgb}{0.1,0.5,0}
\definecolor{goodred}{rgb}{0.7,0,0}
\usepackage[colorlinks,urlcolor=goodgreen,citecolor=blue,linkcolor=goodred]{hyperref}
\usepackage{cleveref}

\makeatletter

\usepackage{soul}

\usepackage{braket}

\usepackage[normalem]{ulem}

\newcommand{\beq}{\begin{equation}}
\newcommand{\eeq}{\end{equation}}
\newcommand{\bea}{\begin{eqnarray}}
\newcommand\bal{\begin{aligned}}
\newcommand\eal{\end{aligned}}
\newcommand{\eea}{\end{eqnarray}}

\makeatother
\begin{document}
\title{Weyl Geometry in Weyl Semimetals}

\author{Giandomenico Palumbo}
\email{giandomenico.palumbo@gmail.com}
\affiliation{School of Theoretical Physics, Dublin Institute for Advanced Studies, 10 Burlington Road, Dublin 4, Ireland}

\begin{abstract}
\noindent  A novel oscillatory behaviour of the DC conductivity in Weyl semimetals with vacancies has recently been identified \cite{Lopes1}, occurring in the absence of external magnetic fields. Here, we argue that this effect has a geometric interpretation in terms of a magnetic-like field induced by an emergent Weyl connection. This geometric gauge field is related to the non-metricity of the underlying effective geometry, which is physically induced by vacancies in the lattice system. Finally, we postulate that the chiral magnetic effect in Weyl semimetals can be affected by the presence of dynamical vacancies.
\end{abstract}
\date{\today}
\maketitle

\noindent \section*{\bf{Introduction}}
\noindent Geometry plays a central role in the physical characterization of various phenomena in condensed matter systems \cite{Volovik}. In the case of topological phases, different geometric models \cite{Volovik2,Cappelli2,Qi2,Abanov,Gromov, Bradlyn1,Bradlyn2,Ye,Palumbo12,Bertolini,Palumbo13,Karabali,Palumbo14} have been formulated to describe these quantum states of matter and corresponding physical features. Thus, the geometric approach complements the gauge theory framework, which involves external and emergent gauge fields allowing us to derive quantum anomalies \cite{Arouca,Semenoff,Ryu,Witten2,Chernodub3,Burrello, Dantas,Palumbo6,Astaneh,Palumbo7}, dualities \cite{Son,Senthil, Vishwanath,Witten,Palumbo4,Palumbo5} and topological quantum field theories \cite{Qi,Essin,Fradkin,Palumbo8,Maggiore,Palumbo9,Palumbo18,Tiwari,Palumbo10,Yu,Palumbo11} to study, for instance, quantum transport and bulk-edge correspondence in topological matter.
In three-dimensional Weyl semimetals (WSMs) \cite{Murakami,Wan,Burkov, Grushin,Armitage,Lv,Hasan,Hasan2,Bouhon}, which are the primary systems under investigation in this work, there have been predicted and observed several quantum effects induced by non-trivial underlying geometries. For instance, strain can induce oscillations in DC conductivity \cite{Franz,Pikulin} and a pseudo-chiral magnetic effect \cite{Pikulin2,Cortijo,Cortijo2,Ilan}. The latter, differently from the standard chiral magnetic effect \cite{Fukushima,Zyuzin,Vazifeh,Parameswaran} induced by the chiral anomaly \cite{Nielsen,Fujikawa,Zumino1,Zumino2}, does not necessarily require the presence of electric and magnetic fields. Strain indeed gives rise to an axial gauge field and corresponding pseudomagnetic field that mimics several effects induced by standard magnetic fields such as the Shubnikov-de Haas effect \cite{Mirlin,Agarwal2} and Landau levels \cite{Behrends}.
 Additionally, other effects may arise from a curved background geometry such as the gravitational chiral anomaly \cite{Eguchi,Witten4,Callan,Landsteiner1,Landsteiner2,Chernodub,Palumbo,Lucas,Gooth,Agarwal,Flachi,Cappelli3,Coriano} and from lattice dislocations, which can lead to a torsionful background geometry with a torsional anomaly and corresponding torsional chiral magnetic effect \cite{Obukhov,Hughes1,Hughes2,Kimura,Fujimoto,Palumbo2,Palumbo3,Ferreiros,Huang,Nissinen1,Nissinen2,Stone,Chernodub2, Chu}.\\
Interestingly, a novel oscillatory behaviour of DC conductivity has been observed in Weyl semimetals when only point defects (vacancies) are present \cite{Lopes1,Lopes2}. This effect cannot be explained neither by strain nor by magnetic fields due to their absence.
In fact, differently from (linear) dislocations, vacancies induce quasi-localized states, which are critical for understanding the oscillatory behaviour of DC conductivity. These states can be interpreted as resonances within a modified geometric framework, where their spatial distribution and energy levels are influenced by the local geometry created by vacancies. As charge carriers traverse these regions, their paths are affected by the geometric configuration of vacancies, leading to constructive and destructive interference patterns that manifest as oscillations in conductivity.\\
Thus, a natural question arises: can we formulate an effective differential-geometric explanation of these quantum oscillations in the continuum low-energy limit?\\
We remind here that in addition to torsion and curvature tensors, generalized differential geometries, known as metric-affine geometries  \cite{Hehl,Eisenhart,Schouten}, are also characterized by the non-metricity tensor, which has been recently employed in an extensive way in quantum field theory in curved spacetime \cite{Janssen,Koivisto2,Zell,Bahamonde}, gravity \cite{Koivisto,Vitagliano,Zanusso,Meyer} and cosmology \cite{Pekar,Iosifidis,Saridakis,Frusciante}.  Non-metricity refers to the failure of an affine connection to preserve the metric tensor during parallel transport. In other words, non-metricity quantifies how much lengths and angles deviate from being preserved as vectors are transported across space.
A particular kind of (torsionless) affine connection, known as Weyl connection \cite{Folland,Ghilencea,Oancea} allows for local rescalings of distances while preserving angles, differently from a Levi-Civita connection that preserves both distances and angles during parallel transport.\\
Inspired by Refs \cite{Gunther,Grachev,Kroner,Katanaev,Clayton,Yavari,Gupta} that relates non-metricity to point defects in lattice systems, here, we employ this geometric approach in the continuum to explain the appearance of the anomalous DC oscillations observed in  WSMs with vacancies. Finally, as a corollary of our geometric theory, we postulate a possible vacancies-induced chiral anomaly and related chiral magnetic effect.

\section*{Weyl connection and fermions}

Here, we discuss the main features of Dirac fermions in the context of Weyl geometry.
In metric-affine gravity,  the metric tensor $g_{\mu\nu}$ and affine connection $\Gamma^\lambda_{\mu\nu}$ are treated as independent variables. Moreover, the affine connection can be always decomposed as follows \cite{Hehl,Eisenhart,Schouten}
\begin{equation}
	{\Gamma}^\lambda_{\mu\nu} = \stackrel{\circ}{\Gamma}{}^\lambda_{\mu\nu} + K^\lambda_{\mu\nu} + L^\lambda_{\mu\nu},
\end{equation}
where $\stackrel{\circ}{\Gamma}{}^\lambda_{\mu\nu}$ is the Levi-Civita connection in Riemannian geometry, 
  $K^\lambda_{\mu\nu}$ is the contortion tensor, encoding torsion $T^\lambda_{\mu\nu} = 2\Gamma^\lambda_{[\mu\nu]}$ and 
$L^\lambda_{\mu\nu}$ is the disformation tensor, encoding non-metricity $Q_{\lambda\mu\nu} = \nabla_\lambda g_{\mu\nu}$, where ${\nabla}_\lambda$ is the covariant derivative associated to
${\Gamma}^\lambda_{\mu\nu}$.\\
Weyl geometry represents a special case of torsionless metric-affine geometry, in which the affine connection is determined by the metric \(g_{\mu\nu}\) and the Weyl connection  (Weyl vector field) \(W_\mu\), which defines the non-metricity tensor \cite{Folland,Ghilencea,Oancea}
\begin{eqnarray}
Q_{\lambda\mu\nu} = {\nabla}_\lambda g_{\mu\nu} = - W_\lambda g_{\mu\nu},
\end{eqnarray}
which transforms in a covariant way under the following transformations
\begin{eqnarray}
	g_{\mu\nu} \to \tilde{g}_{\mu\nu} = e^{\Lambda(x)} g_{\mu\nu}, \nonumber \\
	W_\lambda \to \tilde{W}_\lambda = W_\lambda + \partial_\lambda \Lambda,
\end{eqnarray}
where the rescaling of the metric with respect a spacetime function $\lambda(x)$ is known as conformal transformation of the metric.
Its transformation ensures that it remains compatible with the rescaled metric and transformed Weyl connection, maintaining its role in describing scale-invariant parallel transport.
By taking the trace of the non-metricity tensor, namely $Q_{\mu \nu}{}^\nu = g^{\nu\lambda}Q_{\mu \nu \lambda}$, we have that
\begin{eqnarray}
	W_\mu = \frac{1}{4} Q_{\mu\nu}^{\ \ \nu},
\end{eqnarray}
where the coefficient on the left hand side depends on the number of spacetime dimensions (fixed to four in our case).
Weyl geometry describes a spacetime where local scale invariance (Weyl symmetry) is preserved, and  \(W_\mu\) mediates transformations of the metric under local rescaling.
More concretely, given a vector $v^{\mu}$ as represented in Fig.1, the non-zero Weyl connection causes the change in the length (inner product) of $v^{\mu}$ 
\begin{eqnarray}
	\delta(g_{\mu\nu} v^{\mu}v^{\nu})=\delta g_{\mu\nu} v_{\mu}v_{\nu}=
	-|v|^2 W_{\lambda} dx^{\lambda},
\end{eqnarray}
with $|v|^2=v_\mu v^{\mu}$ the squared length of the vector. 
Thus, a Weyl connection preserves angles but not the distances. This is in contrast to the Levi-Civita connection in Riemannian geometry, which preserves both angles and distances (i.e., it is metric-compatible).
When the Riemannian curvature is negligible, the affine spin connection is determined entirely by $W_\mu$, which can be seen as an effective electromagnetic potential. Moreover, by employing the third Bianchi identity \cite{Schouten}, we can see that the generalized curvature in Weyl geometry is completely encoded by the metric tensor and Weyl connection
\begin{eqnarray}
	R_{(\alpha\beta)\mu\nu}=g_{\alpha\beta}F_{\mu\nu}^W,
\end{eqnarray}
where the round parentheses on the first two indices represent symmetrization and 
\begin{eqnarray}
 F_{\mu\nu}^W=\partial_\mu W_\nu -\partial_\nu W_\mu,
\end{eqnarray}
is the field strength of the Weyl connection.
Notice that $R_{(\alpha\beta)\mu\nu}$ is not null neither in the flat spacetime regime $g_{\mu}=\eta_{\mu}$ (with $\eta_{\mu\nu}$ the Minkowski metric), which is the case we are going to consider in the next sections. Only for the particular case in which $W_\mu=\partial_\mu \phi$, with $\phi$ a scalar field (i.e. the Weyl connection is flat), then $R_{(\alpha\beta)\mu\nu}$ is null (flat Weyl manifolds) and shares the same symmetries of Riemann curvature tensor.\\
\begin{figure}[http]\centering
	\includegraphics[width=8.5cm]{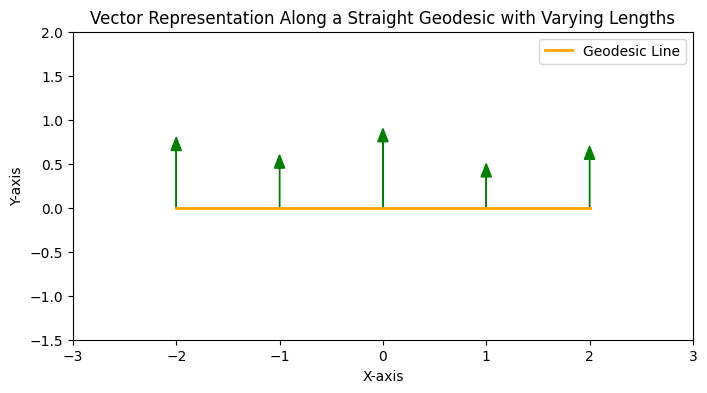}
	\caption{In Weyl geometry, the length of a vector (depicted as a green arrow) is not conserved during parallel transport along a geodesic.}
	\label{BDmode}
\end{figure}

\noindent The coupling between Dirac fermions and Weyl geometry 
requires a careful analysis, because Lorentz-invariant Dirac spinors do not have a minimal coupling with the Weyl connection \cite{Koivisto2}, although there have been proposed some generalizations by employing the projective transformation \cite{Janssen} and non-minimal couplings \cite{Zell}.
Mathematically, this issue comes from the fact that an affine spin connection is in general non-symmetric in the gauge indices and takes values in the GL(4,R) algebra differently from a Lorentz-invariant Dirac field that is related to the SO(3,1)  algebra. In fact, the Lorentz group is just a subgroup of the GL(4,R) group \cite{Tomboulis,Zanusso2}. Thus, to build a more consistent coupling between fermionic matter and affine geometry, we should consider a spinor that transforms under GL(4,R). However, this is not straightforward because there do not exist finite-dimensional spinorial representations of this group and this would invalidate any possible application of a GL-invariant spinor in condensed matter physics where all the physical spinor fields have always a finite number of components.
However, Weyl geometry is a special kind of metric-affine geometry deeply related to conformal geometry. A Weyl manifold is indeed simply a conformal manifold equipped with an additional structure: a Weyl connection \cite{Leigh}.
For this reason, to study fermions in Weyl geometry, we consider a generalized spinor that transforms under the conformal group SO(4,2). This group has been already employed to build a gauge theory of conformal gravity \cite{Kaku}. Moreover, SO(4,2) has a finite-dimensional spinorial representation via its double covering SU(2,2) (to be more precise, the spinor will then transform under SU(2,2)).
This implies that a conformal Dirac spinor is completely well defined as already discussed in Refs \cite{Sla,Westman}. In this context, a SO(3,1)-invariant tangent bundle of general relativity is replaced by a SO(4,2)-invariant tangent bundle that dictates the coupling between matter and spacetime connections by generalizing the Cartan geometry approach already employed in MacDowell-Mansouri gravity and de Sitter-invariant matter theories \cite{Wise,Westman2,Palumbo17,Palumbo16}.
We first introduce the following conformal spin connection \cite{Horne}
\begin{equation}
	A_\mu= e^a_\mu P_a - \frac{1}{2} \omega_{\mu}^{ab} J_{ab}+\lambda^a_\mu K_a + W_\mu D,
\end{equation}
where $a=\{0,1,2,3\}$ and $P_a$, $J_{ab}$, $K_a$ and D are the generators of translations, Lorentz transformations, special conformal transformations and dilatations, respectively. They satisfy the following commutation relations
\begin{eqnarray}
	[J_{ab},J_{cd}]= \eta_{bc} J_{ad}+ \eta_{ad}J_{bc}-\eta_{ac}J_{bd}-\eta_{bd}J_{ac}, \\ \nonumber 
	[J_{ab},P_c]=\eta_{bc}P_a-\eta_{ac}P_b, \hspace{0.1cm}
	[J_{ab},K_c]=\eta_{bc}K_a-\eta_{ac}K_b, \\ \nonumber 
	[P_a,D]=P_a, \hspace{0.2cm} [K_a,D]=-K_a, \\ \nonumber
	[K_a,P_b]= -2 (\eta_{ab}D+J_{ab}),
\end{eqnarray}
where $\eta_{ab}$ is the Minkowski metric.
Moreover, $e^a_\mu$ is the dreibein related to $g_{\mu\nu}$, $\omega_\mu^{ab}$ is the Levi-Civita spin connection, $\lambda_{\mu}^a$ are the gauge fields associated to $K_a$ while $W_\mu$ is indeed the Weyl connection \cite{Cacciatori}.
In the case of spinors, the irreducible SU(2,2) spinorial representation, given by the combinations of $4 \times 4$ Dirac matrices $\gamma_a$, reads \cite{Kaku}
\begin{eqnarray}
	 P_a=-\frac{1}{2}\gamma_a (1-\gamma_5), \hspace{0.2cm} 
	 J_{ab}=\frac{1}{4}[\gamma_a,\gamma_b], \\ \nonumber
	 K_a=\frac{1}{2}\gamma_a (1+\gamma_5), \hspace{0.2cm}  D=-\frac{1}{2}\gamma_5,
\end{eqnarray}
with $\gamma_5=i \gamma_0 \gamma_x \gamma_y \gamma_z$ the chiral matrix.
In this way, we can build the following conformal covariant derivative
\begin{equation}
\nabla^c_\mu = \partial_\mu - i A_\mu.
\end{equation}
We can now insert it in a generalized Dirac Lagrangian and get
\begin{eqnarray}
	L_\psi =  \, i \bar{\psi} \gamma^\mu \nabla^c_\mu \psi,
\end{eqnarray}
where $\psi$ is a four-component spinor field that transforms under SU(2,2) and $\bar{\psi} = \psi^\dagger \gamma^0$ is the adjoint. Notice that this Lagrangian, differently from that one presented in Ref.\cite{Westman}, is not completely invariant under SU(2,2) because $\gamma^\mu$ transforms in a covariant way only under Lorentz transformations. This symmetry breaking is actually not a serious problem, because our starting point was to define a coupling between matter and the background Weyl geometry, in which all the degrees of freedom of the background are encoded in $g_{\mu}$ and $W_{\mu}$ (i.e. an Abelian gauge field). In other words, the spinor theory needs just to be invariant under SO(3,1) and dilatations, although it can be embedded in a more general conformal-invariant theory \cite{Westman}. Firstly, it is well known that the massless Dirac theory is already invariant under the local rescaling of the metric. Secondly, we can now show that $L_\psi$ is also invariant under the U(1) transformation of the Weyl connection. In particular,  because we are interested to apply this theory to WSMs with no curved background and no torsion, we can then take the flat-spacetime limit of $L_\psi$ with
\begin{eqnarray}
	e^a_\mu \rightarrow \delta^a_{\mu}, \hspace{0.2cm} \omega_{\mu}^{ab}\rightarrow 0, \hspace{0.2cm}
	\lambda^a_\mu \rightarrow \delta^a_{\mu},
\end{eqnarray}
where $\delta^a_{\mu}$ is the Kronecker delta. In this way, only the Weyl connection remains non zero and the above Lagrangian simplifies as follows
\begin{eqnarray}\label{Lagrangian}
	L_\psi =  \, \bar{\psi} \gamma^\mu (i \partial_\mu -\gamma_5 W_\mu+...) \psi.
\end{eqnarray}
Here, for convenience, we have rescaled the Weyl connection and the dots represent the residual constant contributions from $P_a$ and $K_a$ terms.
It is straightforward to see that this Lagrangian is invariant under the following transformations
\begin{align}
W_{\mu} \to W_{\mu} + \partial_\mu \Lambda(x), \hspace{0.8cm} \nonumber \\ \quad \psi \to e^{i \Lambda(x)\gamma_5}\psi, \quad \bar{\psi} \to \bar{\psi}  \,e^{i \Lambda(x)\gamma_5}.
\end{align}
Thus, using conformal spinors instead of Lorentz-invariant ones enables us to define a proper minimal coupling between fermionic matter and the Weyl connection. Later in this work, we will explore the implications of the above action in the context of the possible physical observables in WSMs.

\section*{Non-metricity from point defects}

\noindent In this section, we discuss the interesting relation between non-metricity and point defects in crystalline solids. 
In our case, point defects are dilute static vacancies in Weyl semimetals and represented by red dot points in Fig.2. 
In general, a vacancy can be represented by cutting out a small spherical region (a ball) from the Euclidean space which corresponds to removing a portion of the material. After this removal, the boundary of the sphere is shrunk to a single point, leaving behind a "missing" region in the crystal lattice. This process introduces a localized defect in the material, which can be characterized geometrically.
However, differently from linear dislocations, the Burgers vector is absent for point vacancies \cite{Katanaev}. This implies that the torsion tensor cannot be employed to describe them (to be more precise, the torsion tensor is non-zero just at the defect point, where it exhibits a delta-function singularity). As we have seen in the previous section, in metric-affine geometry, there exists a further tensor independent from torsion and curvuare known as non-metricity, which is the correct geometric tool to characterize vacancies are shown in Refs \cite{Gunther,Grachev,Kroner,Katanaev,Clayton}. The space-like part of this tensor $Q_{ijk}$ being not metric compatible, quantifies the deviation in standard length measurements in space.
In a regular crystal, this deviation arises naturally in the presence of point defects. When an observer measures the distance between two atoms by counting atomic steps along crystallographic lines, they encounter interruptions if a vacancy appears instead of a regular atom from the perfect crystal. 
Notice that in the special case of rotational-invariant and isotropic distribution of point defects, the corresponding Weyl connection becomes flat \cite{Yavari,Gupta}. For our purposes, we consider static anisotropic (random) distributions such that the $F^W_{ij}$ is non-zero as well as its corresponding magnetic-like and electric-like fields
\begin{eqnarray}
B_W^i = \epsilon^{ijk} F^W_{ij}, \hspace{0.4cm} E^W_i = F^W_{i0}.
\end{eqnarray}
However, the latter is set to zero because the point defects are assumed to be static and $W_0=0$.
However, the electric-like field could emerge at higher temperatures.
In fact, vacancies are thermally activated, meaning their movement becomes significant only by increasing the temperature of the system \cite{Kroner2}. Consequently, due to thermal effects, the lattice can acquire additional degrees of freedom associated with the motion of vacancies.

\begin{figure}[http]\centering
	\includegraphics[width=8.5cm]{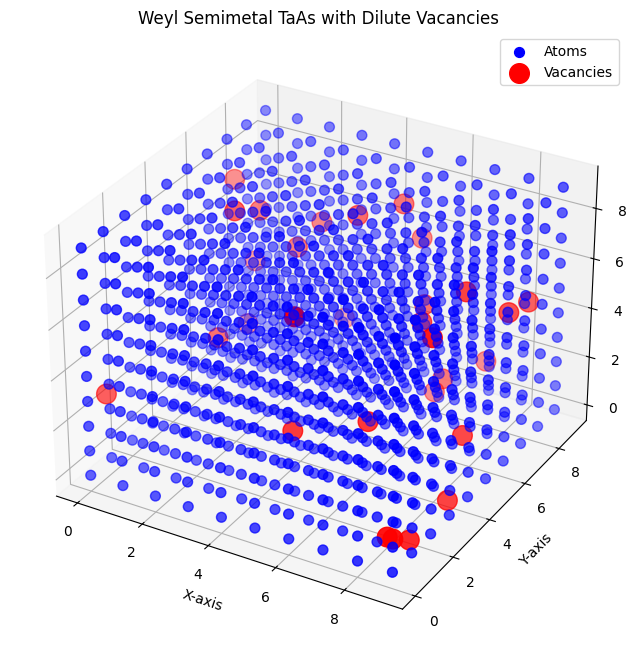}
	\caption{Random distribution of static dilute vacancies (red dots) in a Weyl semimetal TaAs (Tantalum Arsenide), which is depicted by a cubic lattice with lattice sites represented by blue dots.}
\end{figure}

\section*{Implications of the Weyl connection in Weyl semimetals with vacancies}
We are now ready to discuss the main implications of the Weyl geometry in the specific case of time-reversal-invariant Weyl semimetals with dilute vacancies \cite{Lopes1,Lopes2}.  It has been shown in these systems the appearance of unusual oscillation of the DC conductivity even in absence of external magnetic fields. Quantum oscillations in metals and semi-metals are usually associated to the presence of magnetic fields or strain. In fact, the latter can give rise to an axial gauge field and corresponding pseudomagnetic field that generates quantum oscillations \cite{Franz,Pikulin}.
 However, in Refs. \cite{Lopes1,Lopes2}, both magnetic and strain effects are completely absent as well as dislocations that can potentially induce a torsion tensor \cite{Ilan} (torsion couples to relativistic fermions through its axial component, which resembles an axial gauge field).\\
  As already discussed and shown in the previous sections, vacancies can geometrically couple to fermions through an emergent Weyl connection.
 From the Lagrangian in Eq. (\ref{Lagrangian}), we can derive a corresponding momentum-space $4\times 4$ Hamiltonian $H(\mathbf{k},\mathbf{W})$ with the following choice of the Dirac matrices: $\gamma^i=\sigma^x \otimes \sigma^i$, with $i=\{x,y,z\}$, $\gamma^0=i \sigma^y \otimes \sigma^0$, $\gamma^5=-\sigma^z \otimes \sigma^0$, where $ \sigma^i = \{\sigma^x, \sigma^y, \sigma^z\} $ are the Pauli matrices and $\sigma^0$ is the $2 \times 2$ identity matrix, such that
 \begin{eqnarray}
 	H(\mathbf{k},\mathbf{W}) =
 	\begin{pmatrix}
 		H_{+}(\mathbf{k}, \mathbf{W})  & 0 \\
 		0 & H_{-}(\mathbf{k}, \mathbf{W})
 	\end{pmatrix},
  \end{eqnarray}
 with $ \mathbf{k}=(k_x, k_y, k_z) $ the momenta and
 $\mathbf{W}=(W_x, W_y, W_z)$ the space components of the Weyl connection.
The total Hamiltonian basically decomposes into two $2\times 2$ block-diagonal Hamiltonians $H_+$ and $H_-$, which represent the linearized Hamiltonians near the Weyl nodes with opposite chirality
\begin{eqnarray}
 H_{\pm}(\mathbf{k}, \mathbf{W}) = \pm v_F\, \mathbf{\sigma} \cdot (\mathbf{k} \pm \mathbf{W}),
 \end{eqnarray}
 where $ v_F $ is the Fermi velocity (a measure of the linear dispersion slope). Notice that this minimal coupling resembles that one of an axial gauge field induced by strain \cite{Ilan}.
We remind here that  the low-energy limit Hamiltonian of a Weyl semimetal encodes all the relevant topological features of these topological semimetals \cite{Murakami,Wan,Burkov, Grushin,Armitage,Lv,Hasan,Hasan2,Bouhon} and for time-reversal-invariant systems such as TaAs, Weyl cones are at least four, with pairs of nodes related by time-reversal symmetry. For simplicity, we did not add in $H$ the energy offset between the Weyl nodes that breaks inversion symmetry.
Being static, $\mathbf{W}$ can just give rise to magnetic-like effects in our system.  Thus, we can reinterpret the results presented in Ref.~\cite{Lopes1} in a continuum differential-geometric framework, in which the space-dependent and inhomogeneous magnetic-like field $B_W^i$ acts as an effective magnetic-like field and induces oscillations in the DC conductivity.\\
We finally comment on the possible modification of the chiral magnetic effect due to the Weyl connection. In clean topological semimetals in the presence of parallel electric $E_i$ and magnetic $B^i$ fields and non-zero chiral chemical potential $\mu_5$ (the chiral chemical potential is nothing but the difference between the chemical potentials related to the Weyl
fermions with left- and right-handed chiralities), it has been theoretically predicted and experimentally observed the chiral magnetic effect \cite{Fukushima,Zyuzin,Vazifeh,Parameswaran}, which is strictly related to the chiral anomaly originally formulated in quantum field theory \cite{Nielsen,Fujikawa,Zumino1,Zumino2}. The chiral anomaly leads to the non-conservation of the chiral current $j^{5}_\mu=j^L_\mu-j^R_\mu$ (i.e. a difference between left and right currents) which is mathematically expressed as
\begin{eqnarray}
\partial_\mu j^\mu_5 = \frac{e^2}{2\pi^2} E_i B^i,
\end{eqnarray}
where $e$ is the electric charge. This leads to a charge pumping between Weyl nodes of opposite chirality
\begin{eqnarray}
	j^i = \frac{e^2}{2\pi^2} \mu_5 B^i,
\end{eqnarray}
know as chiral magnetic effect.\\
As previously mentioned, the electric-like field $E^W_i$ could be generated by increasing the temperature of the system due to the motion of vacancies \cite{Kroner2}.
In this case, being both $E^W_i$ and $B_W^i$ axial fields and in absence of $B^i$ and $E_i$, the following consistent (chiral) anomaly holds \cite{Arouca}
\begin{eqnarray}
	\partial_\mu j^\mu_5 = \frac{e^2}{6\pi^2}\, E^W_i B_W^i,
\end{eqnarray}
which exactly resembles the consistent anomaly and corresponding chiral magnetic effect induced by strain in WSMs \cite{Ilan,Behrends}.

\section* {\bf Conclusions and outlook}

\noindent Summarizing, we have presented an effective geometric framework in which the oscillatory behaviour of DC conductivity observed in WSMs with point defects can be effectively interpreted through a geometric lens that emphasizes the role of Weyl connections and non-metricity induced by vacancies. By recognizing point defects as effective sources of non-metricity that alters the underlying geometry of electronic states, we gain valuable insights into the mechanisms driving potential novel transport phenomena. In particular, we postulate the modification of the chiral magnetic effect in WSMs in the presence of vacancies. Because the Weyl connection holds for both charged and neutral quasiparticles, we envisage that our results could be generalized to the case of Majorana fermions in Weyl superconductors \cite{Zhang,Burkov2,Meng,Fujimoto2}. Moreover, it would be interesting to analyze the possible geometric interplay of point defects and dislocations. 

\vspace{0.2cm}

\noindent {\bf Acknowledgements:} We thank Saki Koizumi for comments.

\bibliography{references}

\end{document}